\documentclass[12pt,letterpaper]{article}

\usepackage{amsmath}
\usepackage{amsfonts}
\usepackage{amssymb}
\usepackage{rotating}

\newtheorem{theo}{Theorem}
\newtheorem{lem}{Lemma}

\newtheorem{defin}{Definition}
\newtheorem{prop}{Proposition}
\newtheorem{cor}{Corollary}
\newtheorem{rem}{Remark}

\newtheorem{ex}{Example}

\begin{document}

\title{\textsc{Pure Saddle Points and \\ Symmetric Relative Payoff Games\thanks{Some of the material has previously been part of the paper ``Unbeatable imitation'' presented at the International Conference on Game Theory in Stony Brook, 2009.}}}
\author{Peter Duersch\thanks{Department of Economics, University of Heidelberg} \and J\"{o}rg Oechssler\thanks{Department of Economics, University of Heidelberg, Email: oechssler@uni-hd.de} \and Burkhard C. Schipper\thanks{Department of Economics, University of California, Davis, Email: bcschipper@ucdavis.edu}}

\date{February 21, 2010}

\maketitle

\begin{abstract} It is well known that the rock-paper-scissors game has no pure saddle point. We show that this holds more generally: A symmetric two-player zero-sum game has a pure saddle point if and only if it is not a generalized rock-paper-scissors game. Moreover, we show that every finite symmetric quasiconcave two-player zero-sum game has a pure saddle point. Further sufficient conditions for existence are provided. We apply our theory to a rich collection of examples by noting that the class of symmetric two-player zero-sum games coincides with the class of relative payoff games associated with symmetric two-player games. This allows us to derive results on the existence of a finite population evolutionary stable strategies.

\noindent \textbf{JEL-Classifications: } C72, C73.
\end{abstract}

\thispagestyle{empty}

\newpage \setcounter{page}{1}\setlength{\baselineskip}{1.5em}

\section{Introduction}

Many zero-sum games do not have a solution without allowing for mixed actions. What is the class of zero-sum games possessing pure saddle points? Some answers to these questions have been given by Shapley (1964) and Radzik (1991). For instance, Shapley (1964) showed that a finite two-player zero-sum game has a pure saddle point if every 2x2 submatrix of the game has a pure saddle point. Radzik (1991) provided conditions using a discrete notion of quasiconcavity/quasiconvexity. He showed for instance that a two-player zero-sum game whose columns are quasiconcave (i.e. single-peaked) and whose rows are quasiconvex has a pure saddle point if and only if every submatrix ``along the diagonal'' has a pure saddle point. Although both Shapley's and Radzik's results apply to symmetric two-player zero-sum games, none of their results exploits the symmetry property.

In this paper we are interested in pure saddle points of \emph{symmetric} two-player zero-sum games. It is well known that for instance the rock-paper-scissors game has no pure saddle point. We show that this holds more generally. We say that a symmetric two-player game is a \emph{generalized rock-paper-scissors game} if for each column there exists a row with a strictly positive payoff. This notion allows us to characterize symmetric zero-sum games possessing pure saddle points. A symmetric two-player zero-sum game has a pure saddle point if and only if it is not a generalized rock-paper-scissors game. Moreover, we show that every finite symmetric quasiconcave two-player zero-sum game has a pure saddle point. We also provide alternative sufficient conditions for existence in terms of increasing and decreasing differences, potentials and additive separability of payoffs. It turns out that a symmetric two-player zero-sum game has increasing differences if and only if it has decreasing differences. This implies that the payoff function is a valuation. By Topkis (1998) this is equivalent to additively separable payoffs and by Br\^{a}nzei, Mallozzi, and Tijs (2003) it is equivalent to being an exact potential game \`{a} la Monderer and Shapley (1996).

We apply our theory to a rich class of examples. Zero-sum games arise naturally when relative payoffs of arbitrary two-player games are considered.\footnote{There is some experimental evidence that players consider not only their absolute payoffs but also relative payoffs. Early experiments include Nydegger and Owen (1974) and Roth and Malouf (1979). More recently, relative payoff concerns have been studied in behavioral economics and experimental economics under the label of ``inequity aversion''.} The class of symmetric two-player zero-sum games coincides with the class of relative payoff games associated with symmetric two-player games. Pure saddle point actions of relative payoff games characterize finite population evolutionary stable strategies in the underlying game. We apply our result to show the existence of finite evolutionary stable strategy in Cournot duopoly, Bertrand duopoly, public goods games, common pool resource games, minimum effort coordination games, synergistic relationships, arms race, Diamond's search, Nash demand game, rent seeking etc.

The notion of finite population evolutionary stable strategy (fESS) has been introduced by Schaffer (1988, 1989). He also noted the relationship between pure saddle points in relative payoff games and finite population evolutionary stable strategy of the underlying game. This relationship between Nash equilibrium and fESS equilibrium has been analyzed for competitive games by Ania (2008) and for ``weakly competitive'' games by Hehenkamp, Possajennikov, and Guse (2010). Versions of fESS have been applied to learning and evolution in Al\'{o}s-Ferrer and Ania (2005), Hehenkamp, Leininger, and Possajennikov (2004), Leininger (2006), Matros, Temzilidis, and Duffy (2009), Possajennikov (2003), Schipper (2003), Tanaka (2000), and Vega-Redondo (1997). Our results developed here are used in a companion paper, Duersch, Oechssler, and Schipper (2010). There we characterize the class of games in which the decision rule ``imitate-the-best'' is not subject to a money pump by any other decision rule.

In the next section we study the existence of pure saddle points in symmetric two-player zero-sum games. In Section 3 we apply our results to relative payoff games and the existence of finite population evolutionary stable strategies.

\section{Existence in Symmetric Zero-Sum Games}

A two-player zero-sum game $(X_1, X_2, \pi_1, \pi_2)$ consists of two players, player 1 and player 2, and for each player $i \in \{1, 2\}$ a set of pure actions $X_i$ and a payoff function $\pi_i: X_1 \times X_2 \longrightarrow \mathbb{R}$ such that $\pi_1 = \pi$ and $\pi_2 = - \pi$. A two-player zero-sum game is symmetric if $X_1 = X_2 = X$ and $\pi(x, y) = - \pi(y, x)$ for $x, y \in X$. That is, the payoff matrices of symmetric zero-sum games are skew-symmetric.\footnote{A square matrix $M$ is skew-symmetric if $M = - M^T$.} We write $(X, \pi)$ for a symmetric two-player zero-sum game. Note that in a symmetric zero-sum game, the payoffs on the main ``diagonal'' must be zero.

\begin{defin} In a symmetric two-player zero-sum game $(X, \pi)$, a pair of actions $(x, y)$ is a pure saddle point if $\pi(x, y) = \max_{x' \in X} \pi(x', y) = \min_{y' \in X} \pi(x, y')$. A pure saddle point $(x, y)$ is symmetric if $x = y$.
\end{defin}

The following observation is well known.

\begin{rem}\label{symmetry} A symmetric two-player zero-sum game $(X, \pi)$ has a pure saddle point if and only if it has a symmetric pure saddle point.
\end{rem}

\noindent \textsc{Proof. } ``$\Rightarrow$'': In a symmetric game, if $(x'', x')$ is a pure saddle point, so is $(x', x'')$. By the interchangeability of saddle points in two-player zero-sum games, we have that also $(x'', x'')$ and $(x', x')$ are pure saddle points. The converse is straight-forward.\hfill $\Box$\\

Note that if a two-player symmetric zero-sum game has a saddle point, then the saddle-point payoff called the value of the game must be zero (von Neumann, 1928, p. 306). Note further, that given a pure saddle point if an action is replaced by another action yielding the same payoff, then the latter is a saddle point action as well. That is, the game is strongly solvable in the sense of Nash (1951, p. 290-291). Such a strong solution exists only if there is a saddle point in pure actions. This makes pure saddle points very attractive.

It is well known that the symmetric zero-sum game ``Rock-Paper-Scissors'' below has no pure saddle point.
\begin{equation*}
\begin{array}{cc}
&
\begin{array}{ccc}
R~ & ~P & ~S
\end{array}
\\
\begin{array}{c}
R \\
P \\
S%
\end{array}
& \left(
\begin{array}{ccc}
0 & -1 & 1 \\
1 & 0 & -1 \\
-1 & 1 & 0%
\end{array}
\right)%
\end{array}
\end{equation*}
We can generalize the Rock-Paper-Scissors game.

\begin{defin}[Generalized Rock-Paper-Scissors Matrix]\label{grps_matrix} A finite symmetric zero-sum game $(X, \pi)$ is a generalized rock-paper-scissors matrix if in each column there exists a row with a strictly positive payoff to player 1 (i.e., the row player).
\end{defin}

This definition allows us to provide a full characterization of pure saddle points in finite symmetric two-player zero-sum games.

\begin{theo}\label{existence_characterization} A finite symmetric two-player zero-sum game $(X, \pi)$ possesses a pure saddle point if and only if it is not a generalized rock-paper-scissors matrix.
\end{theo}

\noindent \textsc{Proof. } ``$\Leftarrow$'': We show the contrapositive, i.e., if there does not exist a pure saddle point of the symmetric two-player zero-sum game $(X, \pi)$, then $(X, \pi)$ is a generalized rock-paper-scissors matrix. By Remark~\ref{symmetry}, a symmetric zero-sum game has a pure saddle point if and only if it has a symmetric pure saddle point, i.e., if and only if a a combination of actions on the main diagonal is a saddle point. Because $(X, \pi)$ is a symmetric zero-sum game, the payoffs on the diagonal must be zero. Thus, if a symmetric pure saddle point fails to exist, then for each column there must be a row for which the payoff is strictly positive to player 1. This implies that $(X, \pi)$ is a generalized rock-paper-scissors matrix.

``$\Rightarrow$'': Again, we show the contrapositive, i.e., if $(X, \pi)$ is a generalized rock-paper-scissors matrix, then there does not exist a pure saddle point. If $(X, \pi)$ is a generalized rock-paper-scissors matrix, then for each $y \in X$ (i.e., each ``column'') there exists $x \in X$ (i.e., a ``row'') for which the payoff of player 1 is strictly positive. Because $(X, \pi)$ is a symmetric two-player zero-sum game, the payoff on the diagonal must be zero. Hence $y$ is not a best response to $y$, and $(y, y)$ is not a saddle point. Thus for any $y \in X$, $(y, y)$ is not a pure saddle point. \hfill $\Box$\\

Note that a symmetric 2x2 zero-sum game cannot be a generalized rock-paper-scissors game. If one of the row player's off-diagonal relative payoffs is $a > 0$, then the other must be $-a$ violating the definition of generalized rock-paper-scissors matrix. ``Matching pennies'' is not a counter-example because it is not symmetric. Thus we have the following corollary:

\begin{cor}\label{2x2} Every symmetric 2x2 zero-sum game possesses a pure saddle point.
\end{cor}

Theorem~\ref{existence_characterization} provides a characterization with a condition that is easy to check. Nevertheless, it may be useful to know whether ``standard'' second-order conditions imply the existence of pure saddle points as well.

One ``standard'' second-order condition is quasiconcavity. It is known that pure strategy Nash equilibria exist if $X$ is a convex and compact subset of the Euclidean space and $\pi$ is continuous (see Debreu, 1952). Yet, this does not imply the existence in finite games because they are not convex. We follow Radzik (1991) in defining the ``discrete'' analogon of quasiconcavity. It is the notion of single-peakedness.

\begin{defin}[Quasiconcave] A finite symmetric two-player zero-sum game $(X, \pi)$ with the symmetric $m \times m$ payoff matrix $\pi =(\pi _{xy})$ is quasiconcave (or single-peaked) if for each $y \in X$, there exists a $k_{y}$ such that
\begin{equation*}
\pi _{1y}\leq \pi _{2y}\leq \ldots \leq \pi _{k_{y}y}\geq \pi
_{k_{y}+1y}\geq \ldots \geq \pi_{my}.
\end{equation*}
\end{defin}
That is, a symmetric zero-sum game is quasiconcave if each column has a single peak. This definition naturally extends the definition of quasiconcave payoff functions on convex real-valued spaces to the case of finite symmetric two-player zero-sum games.

\begin{theo}\label{existence} Every finite quasiconcave symmetric two-player zero-sum game has a pure saddle point.
\end{theo}

\noindent \textsc{Proof. } Let $(X, \pi)$ be a finite quasiconcave symmetric two-player zero-sum game. Since $(X, \pi)$ is finite, there exists an enumeration of actions from $1$ to $m$. For the proof, we proceed by induction on the set of actions.

Recall that a combination of actions $(k, \ell)$ is a pure saddle point of $(X, \pi)$ if $\pi_{k, \ell}$ is the largest element in column $\ell$ and the smallest element in row $k$.

For $m = 1$ the claim is trivial.

Now let $m > 1$ and assume that there exists a pure saddle point of the symmetric upper block payoff matrix $(\pi_{r, c})_{r, c \leq n < m}$. We will show that there exists also a pure saddle point of the symmetric upper block payoff matrix $(\pi_{r, c})_{r, c \leq n + 1 \leq m}$.

Since $(\pi_{r, c})_{r, c \leq n < m}$ has a pure saddle point, by Remark~\ref{symmetry} it has a symmetric pure saddle point. Let $(k, k)$ be the largest symmetric pure saddle point (with respect to the enumeration of $X$) of $(\pi_{r, c})_{r, c \leq n < m}$.

Case A ($\pi_{n+1, k} \leq \pi_{k, k}$): We claim that if $\pi_{n+1, k} \leq \pi_{k, k}$ then $(k, k)$ is a symmetric pure saddle point of $(\pi_{r, c})_{r, c \leq n + 1 \leq m}$. To see this, note that since $(X, \pi)$ is a symmetric two-player zero-sum game, we must have $\pi_{k, k} = 0$, and $\pi_{k, k} \leq \pi_{k, n+1}$. Hence, $\pi_{k, k}$ remains a largest element in column $k$ and a smallest element in row $k$ after adjoining row $n + 1$ and column $n + 1$ to $(\pi_{r, c})_{r, c \leq n < m}$. Thus $(k, k)$ is a symmetric pure saddle point of $(\pi_{r, c})_{r, c \leq n + 1 \leq m}$.

Case B ($\pi_{n+1, k} > \pi_{n, n}$ and $n = k$): Next we claim that if $n = k$ (i.e., $(n, n)$ is the largest symmetric pure saddle point of $(\pi_{r, c})_{r, c \leq n < m}$) with respect to the enumeration of actions and $\pi_{n+1, n} > \pi_{n, n}$, then $(n+1, n+1)$ is a pure saddle point of $(\pi_{r, c})_{r, c \leq n + 1 \leq m}$. To see this consider the payoff matrix given in Equation~(\ref{matrix1}):
\begin{eqnarray}\label{matrix1} \pi = \left(\begin{array}{ccccc}
& & \vdots & & \vdots \\
\ddots & & \vdots & & \wedge\shortmid (3.) \\
& & \pi_{n, n} & \stackrel{(1.)}{>} & \pi_{n, n+1} \\
& & \wedge &  & \wedge (2.) \\
\cdots & \stackrel{(5.)}{\geq} & \pi_{n+1, n} & \stackrel{(4.)}{>} &  \pi_{n+1, n+1} \\
\end{array}\right)
\end{eqnarray}

We prove all inequalities numbered in the matrix in sequel:
\begin{enumerate}

\item Since $\pi_{n+1, n} > \pi_{n, n}$ and the fact that $(X, \pi)$ is a symmetric two-player zero-sum game, it follows that $\pi_{n, n} > \pi_{n, n+1}$.

\item Since $\pi_{n, n} = \pi_{n+1, n+1} = 0$ and $\pi_{n, n} > \pi_{n, n+1}$, by (1.), we must have $\pi_{n, n+1} < \pi_{n+1, n+1}$.

\item From (2.) and quasiconcavity follows that $\pi_{n, n+1} \geq \pi_{r, n+1}$ for $r \leq n$.

\item From (2.) and the fact that $(X, \pi)$ is a symmetric two-player zero-sum game, it follows that $\pi_{n+1, n} > \pi_{n+1, n+1}$.

\item From (3.) and the fact that $(X, \pi)$ is a symmetric two-player zero-sum game, it follows that $\pi_{n+1, c} \geq \pi_{n+1, n}$ for $c \leq n$.

\end{enumerate} Hence $\pi_{n+1, n+1}$ is the largest element in column $n+1$ and the smallest element in row $n+1$. Thus $(n+1, n+1)$ is a symmetric pure saddle point.

Case C ($\pi_{n+1, k} > \pi_{k, k}$ and $k < n$): Finally, we analyze the case $k < n$ and $\pi_{n+1, k} > \pi_{k, k}$. Consider the payoff matrix given in Equation~(\ref{matrix2}):
\begin{eqnarray}\label{matrix2} \pi = \left(\begin{array}{cccccccccc}
 &        &                       & \vdots     &                     & \vdots       &                     &        &                    & \vdots \\
 & \ddots &                       & \vdots     &                     & \vdots       &                     &        &                    & \wedge\shortmid (4.) \\
 &        &                       & \pi_{k, k}   & \stackrel{(2.)}{=} & \pi_{k, k+1}    & \stackrel{(2.)}{=}  & \cdots & \stackrel{(3.)}{>} & \pi_{k, n+1} \\
 &        &                       & \shortparallel (1.)    &                     & \shortparallel           &                     &        &                    &  \\
 &        &                       & \pi_{k+1, k} & =                   & \pi_{k+1, k+1} &                     & \cdots &                    & \vdots \\
 &        &                       & \shortparallel (1.)    &                     &              &                     &        &                    & \\
 &        &                       & \vdots     &                     &              &                     &        &                    & \vdots \\
 &        &                       & \wedge  &                     &              &                     &        &                    &  \\
 & \cdots & \stackrel{(5.)}{\geq} & \pi_{n+1, k} &                     &              &                     & \cdots &                    & \pi_{n+1, n+1} \\
\end{array}\right)
\end{eqnarray}

\begin{enumerate}

\item We must have $\pi_{r, k} = \pi_{k, k}$ for $n+1 > r \geq k$. To see this, note that if $\pi_{r, k} < \pi_{k, k}$, then we have a contradiction to quasiconcavity. If $\pi_{r, k} > \pi_{k, k}$, then we have a contradiction to $(k, k)$ being a pure saddle point.

\item From (1.) and the fact that $(X, \pi)$ is a symmetric two-player zero-sum game follow that $\pi_{k, k} = \pi_{k, c}$ for $n+1 > c \geq k$.

\item $\pi_{k, n} > \pi_{k, n+1}$ follows from the assumption $\pi_{n+1, k} > \pi_{k, k}$ and the fact that $(X, \pi)$ is a symmetric two-player zero-sum game.

\item Note that by (3.) $\pi_{k, n+1} < 0$. Since $\pi_{n+1, n+1} = 0$, it follows from quasiconcavity that $\pi_{k, n+1} \geq \pi_{r, n+1}$ for $k \geq r \geq 1$.

\item From (4.) and the fact that $(X, \pi)$ is a symmetric two-player zero-sum game follow that $\pi_{n+1, c} \geq \pi_{n+1, k}$ for $c \leq k$.

\end{enumerate}

Since $(k+1, k+1)$ is not a saddle point (because $(k, k)$ is the largest symmetric saddle point with respect to the enumeration of actions by assumption), there must exist a column $c$ such that $\pi_{k+1, c} < \pi_{k+1, k+1}$ with either
\begin{itemize}
\item[(i)] $c < k+1$, or
\item[(ii)] $n+1 > c > k+1$.
\end{itemize}

Consider first case (i). Note that by (5.) in Equation~(\ref{matrix2}) we must have that $\pi_{n+1, c} > \pi_{k+1, c}$. Yet, we also have $\pi_{c, c} = 0 > \pi_{k+1, c}$. These two inequalities contradict quasiconcavity.

Consider now case (ii). Note that by (3.) in Equation~(\ref{matrix2}) we must have that $\pi_{k, c} = 0 > \pi_{k+1, c}$. Yet, we also have $\pi_{c, c} = 0 > \pi_{k+1, c}$. These two inequalities contradict quasiconcavity.

Thus, Case C leads to a contradiction.

Since Cases A, B and C exhaust all cases, we finished the induction step. This completes the proof of the proposition.\hfill $\Box$\\

Note that if the finite zero-sum game is not symmetric but quasiconcave, then it does not need to have a pure saddle point. A counter example is presented in Radzik (1991, p. 26). Hence, symmetry is crucial for the result.

In Example~\ref{grps_qc} below we will show that Theorem~\ref{existence_characterization} is strictly more general than Theorem~\ref{existence}. That is, there are games that are not quasiconcave and not generalized rock-paper-scissors games.

Other ``second-order'' conditions are commonly explored in the literature when analyzing the existence of pure equilibria. We will consider strategic complementarities and substitutes, additive separability, and potentials. Surprising to us, it turns out that for symmetric two-player zero-sum games these conditions are all equivalent.

\begin{defin}[Strategic complementarities and substitutes] Let $X$ be a totally ordered set. A payoff function $\pi$ has decreasing (resp. increasing) differences on $X\times X$ if for all $x^{\prime \prime},x^{\prime },y^{\prime \prime },y^{\prime }\in X$ with $x^{\prime \prime
}>x^{\prime }$ and $y^{\prime \prime }>y^{\prime }$,
\begin{equation*}\label{decreasing diff}
\pi (x^{\prime \prime },y^{\prime \prime })-\pi (x^{\prime },y^{\prime
\prime })\leq (\geq )\pi (x^{\prime \prime },y^{\prime })-\pi
(x^{\prime },y^{\prime }).
\end{equation*}
$\pi$ is a valuation if it has both decreasing and increasing
differences.
\end{defin}

\begin{defin}[Additively Separable] We say that a payoff function $\pi$ is additively separable if $\pi(x, y) = f(x) + g(y)$ for some functions $f, g: X \longrightarrow \mathbb{R}$.
\end{defin}

Potential functions are often useful for obtaining results on convergence of learning algorithms to equilibrium, existence of pure equilibrium, and equilibrium selection. The following notion of potential games was introduced by Monderer and Shapley (1996).

\begin{defin}[Exact potential games]\label{exact_pot} The symmetric two-player game $(X, \pi)$ is an exact potential game if there exists an exact potential function $P:X \times X \longrightarrow \mathbb{R}$ such that for all $y \in X$ and all $x, x' \in X$,
\begin{eqnarray*}
\pi(x, y) - \pi(x', y) & = & P(x, y) - P(x', y), \\
\pi(x, y) - \pi(x', y) & = & P(y, x) - P(y, x').
\end{eqnarray*}
\end{defin}

\begin{lem}\label{Topkis} Let $(X, \pi)$ be an arbitrary symmetric two-player zero-sum game. Then the following statements are equivalent:

\begin{itemize}
\item[(i)] There exists a total order on $X$ and $\pi$ has decreasing differences on $X \times X$,

\item[(ii)] there exists a total order on $X$ and $\pi$ has increasing differences on $X\times X$,

\item[(iii)] there exists a total order on $X$ and $\pi$ is a valuation,

\item[(iv)] $\pi$ is additively separable,

\item[(v)] $(X, \pi)$ has an exact potential.
\end{itemize}
\end{lem}

\noindent \textsc{Proof. } Let $X$ be a totally ordered set such that $\pi$ has decreasing differences on $X \times X$ if for all $x^{\prime \prime \prime },x^{\prime \prime },x^{\prime},x\in X $ with $x^{\prime \prime \prime }>x^{\prime }$ and $x^{\prime
\prime }>x$,
\begin{equation*}
\pi (x^{\prime \prime \prime },x^{\prime \prime })-\pi (x^{\prime
},x^{\prime \prime })\leq \pi (x^{\prime \prime \prime },x)- \pi
(x^{\prime },x).
\end{equation*}
Since $(X, \pi)$ is a symmetric two-player zero-sum game, $\pi
(x^{\prime },x)=-\pi (x,x^{\prime })$ for all $x,x^{\prime }\in X$.
Hence, we can rewrite this inequality as
\begin{equation}
-\pi (x^{\prime \prime },x^{\prime \prime \prime }) + \pi (x^{\prime
\prime },x^{\prime })\leq -\pi (x,x^{\prime \prime \prime })+\pi
(x,x^{\prime }).  \label{rewrite}
\end{equation}
Rearranging inequality~(\ref{rewrite}) yields the definition of increasing
differences,
\begin{equation*}
\pi (x^{\prime \prime },x^{\prime })-\pi (x,x^{\prime })\leq \pi
(x^{\prime \prime },x^{\prime \prime \prime })-\pi (x,x^{\prime \prime
\prime }).
\end{equation*}
Hence (i) if and only if (ii). (iii) follows from the equivalence of (i) and (ii).

By Topkis (1998, Theorem 2.6.4.), a function $\pi(x,y)$ is additively separable on $X\times X$ if and only if $\pi(x,y)$ it is a valuation. Thus, (iii) if and only if (iv).

Br\^{a}nzei, Mallozzi and Tijs (2003, Theorem 1) show that a zero-sum game is an exact potential game if and only if it is additively separable. Hence, (iv) if and only if (v).\hfill $\Box$

\begin{prop}\label{weierstrass} Let $(X, \pi)$ be a symmetric two-player zero-sum game for which $X$ is compact and $\pi$ is upper semicontinuous.
\begin{itemize}
\item[(i)] If $X$ is totally ordered and $\pi$ has decreasing differences on $X \times X$, or

\item[(ii)] if $X$ is totally ordered and $\pi$ has increasing differences on $X\times X$, or

\item[(iii)] if $X$ is totally ordered and $\pi$ is a valuation, or

\item[(iv)] if $\pi$ is additively separable, or

\item[(v)] if $(X, \pi)$ is an exact potential game,

\end{itemize} then a pure saddle point exists. Moreover, for each player, the pure saddle point action is optimal against any of the opponent's actions.
\end{prop}

\noindent \textsc{Proof. } Since $X$ is compact and $\pi$ is upper semicontinuous, any player's best reply correspondence of $(X, \pi)$ is nonempty by Weierstrass' Theorem. Since $\pi$ is additively separable under any property (i) to (v) by Lemma~\ref{Topkis}, the best reply correspondence is constant. Thus, a pure saddle point of $(X, \pi)$ exists.\hfill $\Box$\\

For the remainder of this section, we consider the relationships between the results. Proposition~\ref{weierstrass} is implied by Theorem~\ref{existence} if finite games are considered.

\begin{rem} Let $(X, \pi)$ be a finite symmetric two-player zero-sum game.
\begin{itemize}
\item[(i)] If $X$ is totally ordered and $\pi$ has decreasing differences on $X \times X$, or

\item[(ii)] if $X$ is totally ordered and $\pi$ has increasing differences on $X\times X$, or

\item[(iii)] if $X$ is totally ordered and $\pi$ is a valuation, or

\item[(iv)] if $\pi$ is additively separable, or

\item[(v)] if $(X, \pi)$ is an exact potential game,

\end{itemize} then $(X, \pi)$ is quasiconcave.
\end{rem}

\noindent \textsc{Proof. } By Lemma~\ref{Topkis}, property (iv) holds if and only if any of the other properties hold. If property (iv) holds then there are some functions $f, g: X \longrightarrow \mathbb{R}$ such that $\pi(x, y) = f(x) + g(y)$ for all $x, y \in X$. Since $X$ is finite, by standard utility theory there exists a complete, reflexive, and transitive binary relation $\geq$ on $X$ such that $x' \geq x$ if and only if $f(x') \geq f(x)$. Thus we can order $X$ with respect to $\geq$. Note that this order is independent of $y$. Moreover, note that with $X$ ordered by $\geq$, the game is quasiconcave. \hfill $\Box$\\

Example~\ref{Petergame} in Section~\ref{examples} shows that Theorem~\ref{existence} is not implied by Proposition~\ref{weierstrass}. That is, there are symmetric two-player zero-sum games that are quasiconcave but whose payoff function is not additively separable, a valuation, an exact potential, nor possesses increasing or decreasing differences.

Theorem~\ref{existence} and Proposition~\ref{weierstrass} overlap in the important case of 2x2 games. It is straight-forward to check the following observation:

\begin{rem} Every symmetric 2x2 zero-sum game is quasiconcave, additively separable, a valuation, has increasing and decreasing differences, and has an exact potential.
\end{rem}

The following example demonstrates that Theorem~\ref{existence_characterization} is a strict generalization of Theorem~\ref{existence}.

\begin{ex}\label{grps_qc} Consider the following ``Rock-Paper-Scissors'' game augmented by an additional action ``$B$''.
\begin{equation*}
\begin{array}{cc}
&
\begin{array}{cccc}
R~ & ~P & ~S & ~B
\end{array}
\\
\begin{array}{c}
R \\
P \\
S \\
B
\end{array}
& \left(
\begin{array}{cccc}
0 & -1 & 1 & -1\\
1 & 0 & -1 & -1\\
-1 & 1 & 0 & -1 \\
1 & 1 & 1 & 0
\end{array}
\right)
\end{array}
\end{equation*}
Clearly, it is not a generalized rock-paper-scissors game since for column ``$B$'' there fails to exist a row yielding a strictly positive payoff. Thus, the game possesses a pure saddle point, $(B, B)$. Yet, no matter how actions are ordered, the game fails to be quasiconcave. Hence, there are symmetric two-player zero-sum games that are neither generalized rock-paper-scissors games nor quasiconcave.
\end{ex}

\begin{figure}[h!]
\begin{center}
\includegraphics[scale=.3]{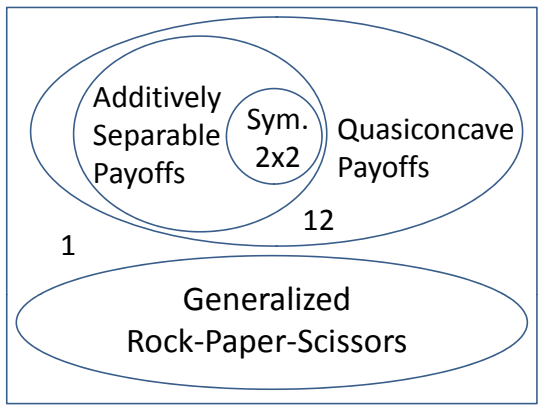}\caption{Relationships among Classes of Finite Two-Player Symmetric Zero-Sum Games\label{relation}}
\end{center}
\end{figure}

Figure~\ref{relation} illustrates the relationships between various classes of games. The numbers refer to the examples.

\section{Applications to Relative Payoff Games}

\subsection{Relative Payoff Games}

Consider now more generally the class of symmetric two-player (not necessarily zero-sum) games $(X, \pi)$ with a (finite or infinite) set of pure actions $X$ and a symmetric and bounded payoff function $\pi :X \times X \longrightarrow \mathbb{R}$. We denote by $\pi(x, y)$ the payoff to the player choosing the first argument. We do not restrict ourselves to zero-sum games.

When instead of the payoff function $\pi$ the relative payoffs are considered, then symmetric two-player games give naturally rise to the class of symmetric zero-sum games.

\begin{defin}[Relative payoff game]\label{zerorel} Given a symmetric two-player game $(X, \pi)$, the associated relative payoff game is $(X, \Delta)$, where the relative payoff function $\Delta: X \times X \longrightarrow \mathbb{R}$ is defined by
\begin{equation*}
\Delta (x, y) = \pi(x, y) - \pi(y, x).
\end{equation*}
\end{defin}
The relative payoff of a player is the difference between his payoff and the payoff of his opponent.

\begin{rem}\label{relzerosum} Every relative payoff game is a symmetric zero-sum game. Conversely, for every symmetric zero-sum game, there is a symmetric two-player game for which the relative payoff game is the symmetric zero-sum game.
\end{rem}

\noindent \textsc{Proof. } Note that by definition, $\Delta (x,y) = -\Delta (y,x)$ and hence $(X, \Delta)$ is a symmetric zero-sum game. For the converse, if $(X, \Delta)$ is a symmetric zero-sum game, then $(X, \frac{1}{2} \Delta)$ is a symmetric two-player game for which $(X, \Delta)$ is the relative payoff game. To see this, note that since $(X, \Delta)$ is a symmetric zero-sum game, we must have that $(X, \frac{1}{2} \Delta)$ is a symmetric zero-sum game. Note further that $\Delta(x, y) = \frac{1}{2} \Delta(x, y) - \frac{1}{2} \Delta(y, x) = \frac{1}{2} \Delta(x, y) + \frac{1}{2} \Delta(x, y)$, where the last equality follows from the fact that $(X, \frac{1}{2} \Delta)$ is a symmetric zero-sum game. \hfill $\Box$\\

The remark shows that every relative payoff game is a symmetric zero-sum game, and that relative payoff games do not impose any restriction on the class of symmetric zero-sum games. Every symmetric zero-sum game is a relative payoff game of some symmetric two-player game. Note also that different symmetric two-player games may have the same relative payoff game.

There is, however, an important conceptual difference between relative payoff games and zero-sum games in general. In relative payoff games, players are assumed by definition to make interpersonal comparisons. No such an assumption is made for general zero-sum games.

\subsection{Finite Population Evolutionary Stable Strategy}

What outcomes in a symmetric two-player game correspond to pure saddle points in its associated relative payoff game? To answer this question we introduce the notion of finite population evolutionary stable strategy (Schaffer, 1988, 1989). A finite population evolutionary stable strategy is the finite population analogue of a neutrally stable strategy (NSS). Different from ``standard'' evolutionary game theory, it is assumed that the entire (finite) population plays together in the same game.

\begin{defin}[fESS] \label{fESS} An action $x^* \in X$ is a \emph{finite population evolutionary stable strategy (fESS)} of the game $(X, \pi)$ if
\begin{equation}\label{fESSformula} \pi(x^*, x) \geq \pi(x, x^*) \mbox{ for all } x \in X.
\end{equation}
\end{defin}

In terms of the associated relative payoff game, inequality~(\ref{fESSformula}) is equivalent to
\begin{equation*}
\Delta (x^*, x) \geq 0 \mbox{ for all } x \in X.
\end{equation*}

Already Schaffer (1988, 1989) observed the relationship between fESS and Nash equilibrium of the relative payoff game.

\begin{prop}[Schaffer, 1988, 1989]\label{Schaffer}Let $(X,\Delta )$ be the relative payoff game derived from the symmetric game $(X,\pi )$ by setting $\Delta (x,y)=\pi (x,y)-\pi (y,x)$. Then the following statements are equivalent:

\begin{itemize}
\item[(i)] $x^*$ is a fESS of $(X,\pi)$,

\item[(ii)] $x^*$ is a pure saddle point action of $(X, \Delta)$.
\end{itemize}
\end{prop}

\noindent \textsc{Proof. } $x^*$ is a symmetric Nash equilibrium action of $(X, \Delta)$ if
\begin{equation*}
\Delta (x^*, x^*) \geq \Delta (x, x^*)\mbox{ for all } x\in X.
\end{equation*}
By symmetry of payoffs and the zero-sum property,
\begin{equation*}
\Delta (x, x^*) = -\Delta (x^*, x).
\end{equation*}
Hence, the inequality is equivalent to
\begin{equation*}
\Delta (x^*, x^*) + \Delta (x^*, x) \geq 0.
\end{equation*}
Since $\Delta (x^*, x^*) = 0$ by definition, we have
\begin{equation*}
\Delta (x^*, x) \geq 0
\end{equation*}
which is precisely the definition of fESS.

Since $(X, \Delta)$ is a symmetric two-player zero-sum game, an action is a Nash
equilibrium action if and only if it a saddle point action. \hfill $\Box$\\

This observation is closely related to Ania (2008) and Hehenkamp, Possajennikov, and Guse (2010) who show for which classes of games fESS and Nash equilibrium coincide.

Uniqueness of the fESS does not matter in the sense that all fESS are relative payoff equivalent. An action $x$ is payoff equivalent to an action $y$ if the payoff from playing $x$ equals the payoff from playing $y$ for every action of the opponent.

\begin{rem} For every symmetric game, all fESS are relative payoff equivalent.
\end{rem}

\noindent \textsc{Proof. } This follows from Proposition~\ref{Schaffer} and the fact that pure saddle points are strong solutions in the sense of Nash (1951, p. 290-291). \hfill $\Box$\\

Theorem~\ref{existence_characterization}, Proposition~\ref{Schaffer}, and Remark~\ref{relzerosum} allow us to provide a characterization of symmetric two-player games who possess a fESS.

\begin{cor} A finite symmetric game $(X, \pi)$ has a fESS if and only if its associated relative payoff game $(X, \Delta)$ is not a generalized rock-paper-scissors matrix.
\end{cor}

From Corollary~\ref{2x2}, Proposition~\ref{Schaffer}, and Remark~\ref{relzerosum} we obtain:

\begin{cor} Every symmetric 2x2 game has a fESS.
\end{cor}

The next two corollaries follows from Theorem~\ref{existence}, Proposition~\ref{Schaffer}, and Remark~\ref{relzerosum}.

\begin{cor}\label{quasiconcave_fESS} Let $(X, \Delta)$ be the relative payoff game associated with a finite symmetric two-player game $(X, \pi)$. If $\Delta$ is quasiconcave, then a fESS of $(X, \pi)$ exists.
\end{cor}

\begin{cor}\label{concaveconvex} Let $(\mathbb{R}^m, \pi)$ be a symmetric two-player game for which $\pi(\cdot, \cdot)$ is concave in its first argument and convex in its second
argument. If the players' actions are restricted to a finite subset $X$ of
the finite dimensional Euclidian space $\mathbb{R}^m$, then a fESS exists.
\end{cor}

Propositions~\ref{weierstrass},~\ref{Schaffer}, and Remark~\ref{relzerosum} imply the following corollary.

\begin{cor}\label{separable} Let $(X, \pi)$ be a symmetric two-player game with $X$ compact and $\pi$ continuous, and the associated relative payoff game be $(X, \Delta)$.
\begin{itemize}
\item[(i)] If $X$ is totally ordered and $\Delta$ has decreasing differences on $X \times X$, or

\item[(ii)] if $X$ is totally ordered and $\Delta$ has increasing differences on $X\times X$, or

\item[(iii)] if $X$ is totally ordered and $\Delta$ is a valuation, or

\item[(iv)] if $\Delta$ is additively separable, or

\item[(v)] if $(X, \Delta)$ is an exact potential game,
\end{itemize} then $(X, \pi)$ has a fESS.
\end{cor}

The following corollary provides a sufficient condition on the payoff function $\pi$ of the underlying game $(X, \pi)$ for conditions (i) to (v) of Corollary~\ref{separable}. We will see in the next section that this condition is satisfied in many well-known textbook examples.

\begin{cor}\label{separable2} Consider a symmetric two-player game $(X, \pi)$ with a compact action set $X$ and a payoff function that can be written as $\pi (x,y)=f(x)+g(y)+a(x,y)$ for some continuous functions $f,g: X \longrightarrow \mathbb{R}$ and a symmetric function $a: X \times X \longrightarrow \mathbb{R}$ (i.e., $a(x,y) = a(y,x)$ for all $x,y \in X$). Then $(X, \pi)$ has a fESS.
\end{cor}

\subsection{Examples\label{examples}}

Existence of fESS in Examples~\ref{linear_Cournot} to~\ref{diamond_search} follows from Corollaries~\ref{separable} or~\ref{separable2}. They demonstrate that the assumption of additively separable relative payoffs is not as restrictive as may be thought at a first glance.

\begin{ex}[Cournot Duopoly with Linear Demand]\label{linear_Cournot} Consider a Cournot duopoly given by the symmetric payoff function by $\pi(x,y)=x(b-x-y)-c(x)$ with $b>0$. Since $\pi (x,y)$ can be written as $\pi(x,y)=bx-bx^{2}-c(x)-xy$, Corollary~\ref{separable2} applies and a fESS exists.
\end{ex}

\begin{ex}[Bertrand Duopoly with Product Differentiation]\label{Bertrand} Consider a differentiated duopoly with constant marginal costs, in which firms 1 and 2 set prices $x$ and $y$, respectively. Firm 1's profit function is given by $\pi (x,y)=(x-c)(a+by-\frac{1}{2}x)$, for $a>0$, $b \in [0,1/2)$. Since $\pi(x, y)$ can be written as $\pi (x,y) = ax-ac+\frac{1}{2}cx-\frac{1}{2}x^{2}-bcy+bxy$, Corollary~\ref{separable2} applies and a fESS exists.
\end{ex}

\begin{ex}[Public Goods]\label{public_goods} Consider the class of symmetric public good games defined by $\pi (x,y)=g(x,y)-c(x)$ where $g(x,y)$ is some symmetric monotone increasing benefit function and $c(x)$ is an increasing cost function. Usually, it is assumed that $g$ is an increasing function of the the sum of provisions, that is the sum $x+y$. Various assumptions on $g$ have been
studied in the literature such as increasing or decreasing returns. In any
case, Corollary \ref{separable2} applies and a fESS exists.
\end{ex}

\begin{ex}[Common Pool Resources]\label{CPR} Consider a common pool resource game with two appropriators. Each appropriator has an endowment $e>0$ that she can invest in an outside activity with marginal payoff $c>0$ or into the common pool resource. $x \in X \subseteq \lbrack 0,e]$ denotes the maximizer's investment into the common
pool resource (likewise $y$ denotes the imitator's investment). The return
from investment into the common pool resource is $\frac{x}{x+y}
(a(x+y)-b(x+y)^{2})$, with $a,b>0$. So the symmetric payoff function is
given by $\pi (x,y)=c(e-x)+\frac{x}{x+y}(a(x+y)-b(x+y)^{2})$ if $x,y>0$ and $ce$ otherwise. (See Walker, Gardner, and Ostrom, 1990.) Since $\Delta
(x,y)=(c(e-x)+ax-bx^{2})-(c(e-y)+ay-by^{2})$, Corollary~\ref{separable}
implies the existence of a fESS.
\end{ex}

\begin{ex}[Minimum Effort Coordination]\label{minimum_effort} Consider the class of minimum effort games given by the symmetric payoff function $\pi (x,y)=\min \{x,y\}-c(x)$ for some cost function $c$ (see Bryant, 1983 and Van Huyck, Battalio, and Beil, 1990). Corollary~\ref{separable2} implies that a fESS exists.
\end{ex}

\begin{ex}[Synergistic Relationship] Consider a synergistic relationship among two individuals. If both devote more effort to the relationship, then they are both better off, but for any given effort of the opponent, the return of the player's effort first increases and then decreases. The symmetric payoff function is given by $\pi(x, y) = x (c + y - x)$ with $c > 0$ and $x, y \in X \subset \mathbb{R}_+$ with $X$ compact (see Osborne, 2004, p. 39). Corollary~\ref{separable2} implies that the existence of fESS.
\end{ex}

\begin{ex}[Arms Race]\label{arms_race} Consider two countries engaged in an arms race (see e.g. Milgrom and Roberts, 1990, p. 1272). Each player chooses a level of arms in a compact totally ordered set $X$. The symmetric payoff function is given by $\pi(x,y)=h(x-y)-c(x)$ where $h$ is a concave function of the difference between
both players' level of arms, $x-y$, satisfying $h(x-y)=-h(y-x)$. Corollary~\ref{separable} implies that a fESS exist.
\end{ex}

\begin{ex}[Diamond's Search]\label{diamond_search} Consider two players who exert effort searching for a trading partner. Any trader's probability of finding another particular trader is proportional to his own effort and the effort by the other. The payoff function is given by $\pi(x, y) = \alpha x y - c(x)$ for $\alpha > 0$ and $c$ increasing. (See Milgrom and Roberts, 1990, p. 1270.) The relative payoff game of this two-player game is additively separable. By Corollary~\ref{separable} implies the existence of a fESS.
\end{ex}

\begin{ex}[Nash Demand Game] Consider the Nash Demand game as follows (see Nash, 1953):\footnote{Interestingly, early experimental evidence for relative payoff concerns were found when testing Nash bargaining (Nydegger and Owen, 1974).} Two players simultaneously demand an amount in $\mathbb{R}_+$. If the sum is within a feasible set, i.e., $x + y \leq s$ for $s > 0$, then player 1 receives the payoff $\pi(x, y) = x$ if $x + y \leq s$, and $\pi(x, y) = 0$ otherwise (and symmetrically for player 2). The relative payoff function is quasiconcave. If the players' demands are restricted to a finite set, then Corollary~\ref{quasiconcave_fESS} implies the existence of fESS.
\end{ex}

\begin{ex}[Rent Seeking]\label{rentseeking} Two contestants compete for a rent $v>0$ by bidding $x,y\in X\subseteq \mathbb{R}_{+}$. A player's probability of winning is
proportional to her bid, $\frac{x}{x+y}$ and zero if both players bid zero.
The cost of bidding equals the bid. The symmetric payoff function is given
by $\pi (x,y)=\frac{x}{x+y}v-x$ (see Tullock, 1980, and Hehenkamp,
Leininger, and Possajennikov, 2004). $\pi(x, y)$ is concave in $x$ and convex in $y$. Thus Corollary~\ref{concaveconvex} implies that a fESS exists.
\end{ex}

\begin{ex}\label{Petergame} Consider a symmetric two-player game with the payoff function given by $\pi (x,y) = \frac{x}{y}$ with $x, y \in X \subset [1, 2]$ with $X$ being finite. This game's relative payoff function is quasiconcave. Corollary~\ref{quasiconcave_fESS} implies the existence of fESS of $(X, \pi)$. Moreover, the example demonstrates that not every quasiconcave relative payoff function is additively separable.
\end{ex}

The following example shows that the pure saddle point of the relative payoff game and the fESS of the original game may not necessary coincide with Nash equilibrium of the underlying game.\footnote{See Ania (2008) and Hehenkamp, Possajennikov, and Guse (2010) for general results.} Moreover, it shows that extremely inefficient outcome may be selected by fESS.

\begin{ex} Consider the symmetric 2x2 game given in the following payoff matrix:
\begin{equation*}
\begin{array}{cc}
&
\begin{array}{cc}
\mbox{ A } & \mbox{ B }%
\end{array}
\\
\begin{array}{c}
\mbox{A} \\
\mbox{B}%
\end{array}
& \left(
\begin{array}{cc}
4,4 & 1,2 \\
2,1 & 0,0%
\end{array}%
\right)
\end{array}%
\end{equation*}%
This game has a unique Nash equilibrium, $(A,A)$, that is efficient and in
strictly dominant actions. Yet, the unique fESS is $(B,B)$.
\end{ex}

\end{document}